\documentclass[11pt]{article}

\usepackage{amsmath,amssymb,amsthm}
\usepackage{graphicx,hyperref}
\usepackage{geometry}
\usepackage{subfigure}
\makeatletter

\newcommand{\Rmnum}[1]{\expandafter\@slowromancap\romannumeral #1@}
\makeatother

\hypersetup{
    bookmarks=true,                             
    unicode=false,                              
    pdftoolbar=true,                            
    pdfmenubar=true,                            
    pdffitwindow=true,                          
    pdftitle={Synchronization},      
    pdfauthor={Kramer,Bollt},  
    pdfsubject={Synchronization},                  
    pdfnewwindow=true,                          
    pdfkeywords={keywords},                     
    colorlinks=true,                            
    linkcolor=blue,                             
    citecolor=blue,                             
    filecolor=magenta,                          
    urlcolor=blue                               
}

\title{Spatially Dependent Parameter Estimation and Nonlinear Data Assimilation by Autosynchronization of a System of Partial Differential Equations}
\author{Sean Kramer and Erik Bollt}
\date{}

\begin{document}
  \maketitle

  \begin{abstract}
    Given multiple images that describe chaotic reaction-diffusion dynamics, parameters of a PDE model are estimated using autosynchronization, where parameters are controlled by synchronization of the model to the observed data. A two-component system of predator-prey reaction-diffusion PDEs is used with spatially dependent parameters to benchmark the methods described. Applications to modelling the ecological habitat of marine plankton blooms by nonlinear data assimilation through remote sensing is discussed. 
  \end{abstract}

  \section{Introduction}\label{sec:intro}
  Parameter estimation in ODEs and PDEs has developed into a vast field in applied mathematics and control engineering. For models representing important physical processes, accurate estimates of appropriate model parameters may help inform short-term management decisions by model forecasting. However, to forecast a system one requires not only accurate parameter estimates, but also full knowledge of the initial state of the system. There are widely varying and powerful methods for parameter estimation of spatio-temporal systems including, but certainly not limited to Kalman filter methods \cite{schiff08,annan05,wan00}, multiple shooting methods \cite{Muller04,muller02}, and adjoint methods \cite{navon98}. A method of parameter estimation based on synchronization has drawn substantial interest  \cite{PC90,P96,PJK96,SKP96,YCCLP07,YP08,SO09,QBCKA09,SSLP10,BLP11,SP11}. Applications include communications and cryptography \cite{SKP96}, electronics and circuit dynamics \cite{PJK96,KTP97}, and cardiac cell dynamics \cite{SSLP10} to name just a few. There are currently several methods to estimate parameters based on synchronization. One approach is to optimize a time-averaged synchronization error on which synchronization acts as a regularizing force; the optimization problem of finding the minimum synchronization error in parameter space is well-posed \cite{PJK96,QBCKA09,SSLP10,BLP11,SP11}. Our interest here will be based on an approach to force a response model to adapt to observed data by developing additional equations for the parameters that depend on the synchronization error \cite{P96,SO09}. 
  
  To estimate model parameters by synchronization, we exploit a special variation of synchronization called ``autosynchronization''. For systems of ODEs, an observed scalar time series is coupled to a response system during model simulation. The goal of this feedback is to cause the response system to synchronize to the drive system. Ideally a proof of convergence follows by demonstration of an appropriate Lyapunov function \cite{P96,YCCLP07}. In \cite{YP08,SO09} we see some generalizations of how to derive synchronization schemes for many systems including the case where, apriori, we do not know the model form of the drive system \cite{SO09}. By autosynchronization, we can recover the model parameters, the current model state, and in some cases, a model form for an observed system. 
  
Stating an autosynchronization problem in the ODE setting, we require a drive system 
\begin{equation} \label{eq:DriveModel}
\mathbf{u_t} = \mathbf{f(u,p)},
\end{equation} 
  from which we are able to sample data with (unknown to us) parameters $\mathbf{p} \in \mathbf{R^m}$. Then we must state a response system 
\begin{equation} \label{eq:ResponseModel}
\mathbf{v_t} = \mathbf{g(u,v,q)}
\end{equation} 
  with the same model form as the drive system if $\mathbf{q = p}$. By ``same'' we mean in as far as possible by our understanding of the underlying physics. Then the goal is that when $\mathbf{u}$ is coupled forward into Eq \eqref{eq:ResponseModel}, then Eq \eqref{eq:ResponseModel} will synchronize with Eq \eqref{eq:DriveModel} and  $\mathbf{u \rightarrow v}$. Furthermore, parameter ODEs are given by
\begin{equation} \label{eq:ParameterODE}
\mathbf{q_t} = \mathbf{h(u,v,q)}
\end{equation} 
so that $\mathbf{(v,q)} \rightarrow \mathbf{(u,p)}$ as $t \rightarrow \infty$.
  
The idea of synchronization was extended to one-dimensional systems of PDEs in \cite{KTP97} and two-dimensional systems in \cite{BLP11}, where the authors considered the Grey-Scott and Barkely reaction-diffusion systems respectively. In these works, the authors observed synchronization of an infinite-dimensional system by coupling the drive and response systems at only a finite number of grid points. Further work examines parameter estimation for given PDE systems using optimization over the synchronization error surface as discussed above. The authors observe single-species assimilation as they drive the PDE system to synchronization while coupling with only one species \cite{BLP11}. However, none of these utilize autosynchronization. 

In many systems, it is very reasonable to expect that model parameters need not be spatially homogeneous. For example, taking our problem of interest, spatial inhomogeneity in parameter values may be central when constructing models for coastal algal blooms, since plankton growthrate is affected by near-shore nutrient runoff and upwelling \cite{M02}. More to that point, ocean fronts and eddies cause flow-induced long-term inhomogeneities in the ocean which results in a formidable spatial structure for density profiles in the ocean \cite{M02}. Whether inhomogeneities be the result of the flow dynamics or of ``boundary conditions" from nutrient runoff, they are an important consideration for modelling ecology over large coastal domains. Thus it is reasonable to argue that a biophysics-based model should accept spatially dependent parameters. 

A drawback to the aforementioned methods, including Kalman filter and multiple shooting methods, is that they do not consider spatially dependent parameter values, a priority noted in \cite{schiff08}. Parameter estimation by filtering methods adapted for PDEs can be computationally expensive \cite{muller02}. Furthermore, we have found that some filtering methods have trouble during periods of exponential growth, such as might be expected during plankton blooms. Optimizing the time-averaged synchronization error in some function space is far more complicated than the finite-dimensional alternative with scalar parameters as in \cite{BLP11}; optimization methods may not be practical. 
  
Our work aims to extend the method of parameter estimation for PDE systems by synchronization to autosynchronization, especially including autosynchronization with spatially dependent parameters.  Thus, we investigate observed data from the PDE drive system
\begin{equation} \label{eq:DriveModelPDE}
\mathbf{u_t}(x,y,t) = \mathbf{f(u}(x,y),\mathbf{p}(x,y))
\end{equation} 
with parameters $\mathbf{p}(x,y) \in C^0(\Omega)$ and a response system 
\begin{equation} \label{eq:ResponseModelPDE}
\mathbf{v_t}(x,y,t) = \mathbf{g(u}(x,y),\mathbf{v}(x,y),\mathbf{q}(x,y)).
\end{equation} 
We formulate an associated system of PDEs for the parameters of Eq \eqref{eq:ResponseModelPDE}
\begin{equation} \label{eq:ParameterEqn}
\mathbf{q_t}(x,y,t) = \mathbf{h(u}(x,y),\mathbf{v}(x,y))
\end{equation} 
with the goal that $\mathbf{(v,q)} \rightarrow \mathbf{(u,p)}$ as $t \rightarrow \infty$.
We design our methods considering a benchmark system of reaction-diffusion PDEs. Since we know the model form of the drive system  and the parameters used to build the observed data, we can compare our estimated parameters with the exact parameters. 

In preview of the paper layout, we begin by introducing the reaction-diffusion equations that we will use as the drive system. We discuss how this system is solved numerically and the parameters used to simulate complex spatiotemporal dynamics. Next, we implement the response system and show the parameter PDEs used to find autosynchronization. We demonstrate the power of the systems of PDEs to autosynchronize by employing three different spatial functions for the parameters. Next, we show the estimated parameters and the convergence plots for both state variables and parameters to the correct values. Finally, we give an improvement on the response system that admits autosynchronization wherein only one species is sampled, which is an important breakthrough for applications since generally only the phytoplankton is easily observable.


\section{The Parameter Estimation Method}\label{sec:RD}
Consider the system of two PDEs as given in \cite{M02},
\begin{eqnarray} \label{eq:Fish}
\frac{\partial P}{\partial t} &=& \triangle P + P(1-P) - \frac{P Z}{P + h},\\
\nonumber \frac{\partial Z}{\partial t} &=& \triangle Z + k\frac{P Z}{P+h} - mZ,
\end{eqnarray} 
on a compact connected two-dimensional domain, $\Omega$, with zero-flux boundary conditions. 

In terms of the biology of the model, the system represents a dimensionless reaction-diffusion model for phytoplankton-zooplankton predator prey dynamics in a horizontal layer where vertical distributions of plankton are considered uniform. For simulations shown here, we choose $\Omega$ to be a rectangle of size $864 \times 288$. Although shown in dimensionless form, the model is derived from principles in which phytoplankton concentrations obey a logistic growth and are grazed upon by zooplankton, following a Holling-type \Rmnum{2} functional response. First classified by Holling, \cite{Holling59}, the Holling-type \Rmnum{2} functional response assumes a decelerating growth rate such that the predator, or consumer, is limited by its ability to efficiently process food. Zooplankton grow at a rate, k, proportional to phytoplankton mortality and are subject to a natural mortality rate m. For certain parameters, this system gives rise to transient spiral pattern behaviour on its way to spatially irregular patchy patterns \cite{M02}. In \cite{M02}, parameters are set to $k = 2$, $h = 0.4$, and $m = 0.6$. For homogeneous initial plankton distributions, the system remains in a homogeneous state for all time so we use the perturbed initial conditions found in \cite{M02}. 

We solve this system using a finite difference method, with a nine-point center difference stencil for spatial derivatives and forward Euler time stepping. Our simulations use spatial discretization with $dx = 2$ and and Euler time step of $dt = .2$. The model output is treated as an image sequence given by a particular (known) model form but with \textit{unknown} parameters $k$ and $m$, to be determined. Thus we will mimic our target application of remote sensing oceanographic images of hyperspectral images filtered to reveal plankton blooms. Further, we will be interested to allow $k$ and $m$ to vary spatially as functions, $k(x,y)$ and $m(x,y)$.

Our interest in this PDE model stems from our work  in remote sensing, to build a better understanding of our ocean's ecology. Particularly, we aim to predict short term behavior of coastal algal blooms. Such a system may in principle be modelled by estimating parameters directly from observed data in the field. However, hyperspectral satellite imagery provides the observed data to which we would synchronize a response model in hopes of autosynchronization providing good parameter estimates for forecasting. Since phytoplankton are largely affected by spatial inhomogeneities in the ocean such as nitrogen runoff, regions of hypoxia, or upwelling, to name a few parameter inhomogeneity-inducing effects, we wish to allow model parameters to vary spatially. These considerations are especially important since our models will be built over coastal domains where large changes in ocean biology occur spatially, leading naturally to spatially dynamic parameters. 

We are only able to observe time series data Eq (\ref{eq:DriveModelPDE}) as a movie and we know the model form of Eq (\ref{eq:Fish}), but want to estimate the parameters used to create the observed data. The system of Eq (\ref{eq:Fish}) will be taken as the drive system and we form a response system to be synchronized to the observations as,

\begin{eqnarray}\label{eq:Fishr}
\frac{\partial \hat{P}}{\partial t} &=& \triangle \hat{P} + \hat{P}(1-\hat{P}) - \frac{\hat{P} \hat{Z}}{\hat{P} + h} + \kappa(P-\hat{P}),\\
\nonumber \frac{\partial\hat{Z}}{\partial t} &=& \triangle \hat{Z} + \hat{k}\frac{\hat{P} \hat{Z}}{\hat{P}+h} - \hat{m}\hat{Z} + \kappa(Z-\hat{Z}),
\end{eqnarray} 
where we assume $\hat{P}(x,y,0) \neq P(x,y,0)$, $\hat{Z}(x,y,0) \neq Z(x,y,0)$, $\hat{k}(x,y,0) \neq k(x,y)$, and $\hat{m}(x,y,0) \neq m(x,y)$. Thus, we do not know the initial model states, and wish to recover the spatially varying parameters $m(x,y)$ and $k(x,y)$. To derive Eq (\ref{eq:Fishr}), a diffusive coupling term is added to each equation in Eq (\ref{eq:Fish}) accounting for the error between the drive and response values with a coupling strength, $\kappa$. These additional terms drive $\hat{P} \rightarrow P$ and $\hat{Z} \rightarrow Z$, so that the PDEs will synchronize after a short time. The synchronization is of identical type and dependent upon the choice of $\kappa$, as is the synchronization speed.


\section{Results and Simulations of Autosynchronization Parameter Estimation}
We modify the system Eq (\ref{eq:Fish}) as found in \cite{M02} by allowing the parameters to be nonnegative $C^0(\Omega)$ functions. Here $\Omega$ is the domain, which in the case of our simulations, $\Omega \subset \mathbf{R}^2$ is a compact domain such as a rectangle or even a domain shaped as the Gulf of Mexico. Parameters are updated as diffusively coupled PDEs during the synchronization process as
\begin{eqnarray}\label{eq:ParUpdate}
\frac{\partial \hat{k}}{\partial t} &=& -s(P - \hat{P}) \ \ \ \ s > 0, \\
\nonumber \frac{\partial\hat{m}}{\partial t} &=& -s(Z - \hat{Z}),
\end{eqnarray} 
where we choose $s = 30$ for specificity and for which we observe good convergence results. The parameter equations are evolved simultaneously with Eq \eqref{eq:Fishr} with a forward Euler discretization and the same time step. The model form of Eq \eqref{eq:ParUpdate} was chosen after testing several forms and there may exist other forms for which synchronization is possible. Once the model form was chosen, a parameter search was performed to find $s=30$. As we vary $s$ and $\kappa$, autosynchronization can fail, a common situation with diffusively coupled systems. Parameters may be updated as reaction-diffusion PDEs, by adding a diffusion term, however we need to restrict parameters to be nonnegative $C^2(\Omega)$ functions and stability may be affected. 
To begin the simulation, parameters are initialized as the constant function
\begin{eqnarray}\label{eq:ParInit}
\hat{k}(x,y,0) &=& 10,\\
\hat{m}(x,y,0) &=& 10. 
\end{eqnarray}
We evolve Eq \eqref{eq:Fish} forward and count the model output as observed data. Initial conditions for the response system are $\hat{P}(x,y,0) = 0$ and $\hat{Z}(x,y,0) = 0$. Furthermore, to avoid values outside the normal range of Eq\eqref{eq:Fish}, we enforce that 
\begin{eqnarray}
\nonumber \hat{P} = \Bigg\{ \begin{matrix} \hat{P}&:& 0 < \hat{P} < 2 \\0&:& \hat{P} \leq 0\\2&:& \hat{P} \geq 2 \end{matrix},
\end{eqnarray}
during the simulation. 

\begin{figure}[h]
\begin{center}
\begin{minipage}{3.2in}
\includegraphics[width=3.0in]{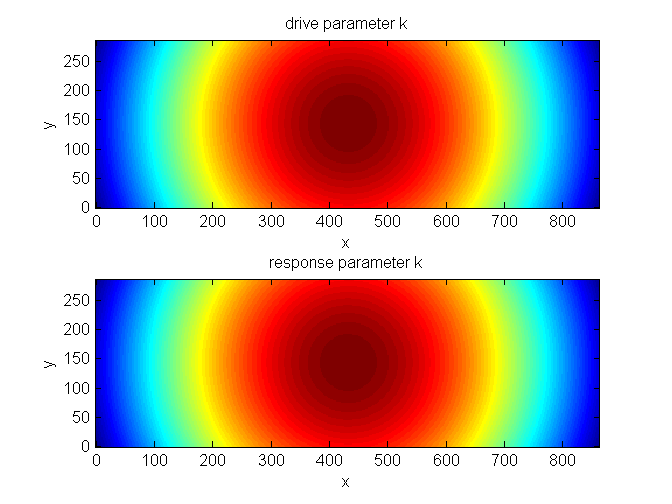}(a)
\end{minipage}
\begin{minipage}{3.2in}
\includegraphics[width=3.0in]{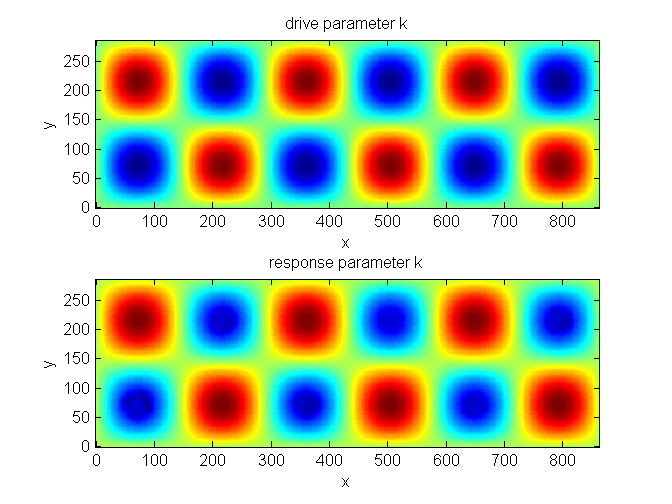}(b)
\end{minipage}
\caption{Autosynchronization of spatially dependent parameters at $t=2000$. Drive (top) vs response (bottom). Left: Model parameters given by $k_2(x,y)$ (shown) and $m_2(x,y)$. Right: Model parameters given by $k_3(x,y)$ (shown) and $m_3(x,y)$.}
\label{fig:figP2}
\end{center}
\end{figure}

First, we develop synthetic datasets with spatially varying parameters to challenge our methods. Spatially dependent parameters are chosen to be in the range given in \cite{M02} for spatially irregular behavior. Three different functional forms for the parameters are tested. First, we use a linearly varying parameter set
\begin{eqnarray}\label{eq:Linear}
k_1(x,y) &=& a \left(\frac{x}{n} + \frac{y}{m} + l\right) + b, \\
\nonumber m_1(x,y) &=& c \left(\frac{x}{n} + \frac{y}{m} + l\right) + d,
\end{eqnarray}
where $m$ and $n$ represent the size of the domain shown in Figure ~\ref{fig: Domain} and $a = 2, b = 0.14, c = 0.6, d = 0.1$, and $l = 5$. Thus, appropriate parameters are chosen to maintain $m(x,y)$ and $k(x,y)$ in the target range. Next, we define a Gaussian parameter function as

\begin{eqnarray}\label{eq:Gauss}
k_2(x,y) &=& a e^{-\left(\frac{(x-n/2)^2}{2\sigma^2} + \frac{(y-m/2)^2}{2\sigma^2}\right)}, \\
\nonumber m_2(x,y) &=& c e^{-\left(\frac{(x-n/2)^2}{2\sigma^2} + \frac{(y-m/2)^2}{2\sigma^2}\right)},
\end{eqnarray}
where $a,b,m,n$ have the same values. For example, $k_2(x,y)$ is shown on the top half of (a) in Figure \ref{fig:figP2}. Finally, we define

\begin{eqnarray}\label{eq:Sineplot}
k_3(x,y) &=& a \ \cos(b x + d)\sin(by) + s, \\
\nonumber m_3(x,y) &=& c \ \cos(b x + d)\sin(by) + t,
\end{eqnarray}
where $a = 0.2, b = \pi/(m/2), c = 0.6, d = \pi/2, s = 0.5$, and $t = 1.5$, to test the quality of the autosynchronization scheme to resolve fine structures in model parameters. The surface produced by $k_3(x,y)$ is displayed on the top half of (b) in Figure \ref{fig:figP2}. We observe solution data at every time step relative to the response system, Eq \eqref{eq:Fishr}, and the parameter system, Eq \eqref{eq:ParUpdate}, to drive $(\hat{P},\hat{Z}) \rightarrow (P,Z)$ and $(\hat{m}(x,y),\hat{k}(x,y)) \rightarrow (m(x,y),k(x,y))$ as $t \rightarrow \infty$. For brevity, only the parameters defined by Eq \eqref{eq:Gauss} and Eq \eqref{eq:Sineplot} are shown and compared with their estimated counterparts.

We observe autosynchronization for each test set of parameters and the spatial inhomogeneities in each case are effectively resolved. 
\begin{figure}[h]
\begin{center}
\subfigure{\includegraphics[width=3.2in]{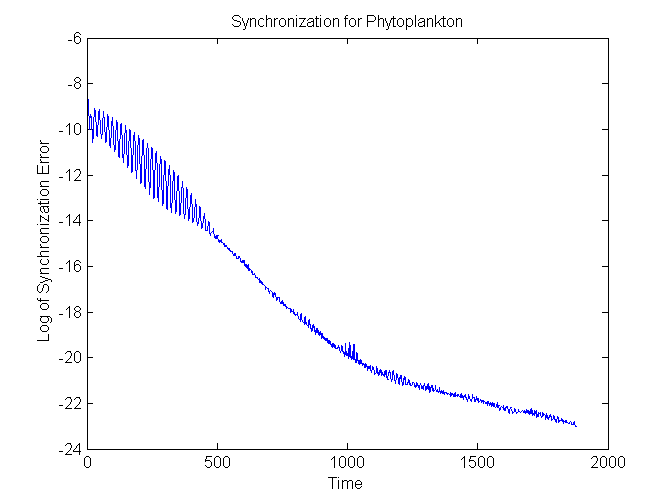}}
\subfigure{\includegraphics[width=3.2in]{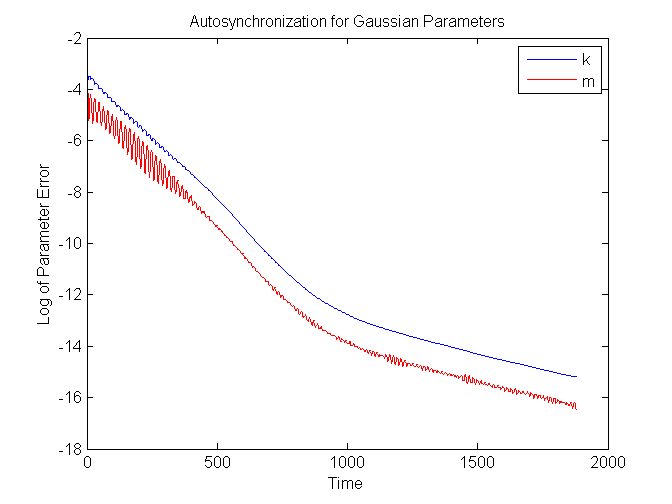}}
\caption{Left: Globally averaged synchronization error between drive and response PDEs (Phytoplankton) on a log scale. Right: Globally averaged synchronization error between drive and response parameters on a log scale. Both plots obtained using model parameters in Eq \eqref{eq:Gauss}.}
\label{fig:figP1}
\end{center}
\end{figure}
In Figure (\ref{fig:figP1}), the globally averaged error between the drive and response PDEs (Phytoplankton density) has been driven to less than $1.0 \times 10^{-10}$ and the globally averaged error between true and estimated parameters has been driven to below $1.0 \times 10^{-6}$. Interestingly, the results show a change of convergence rate midway through the simulation. For these results, we choose $\kappa = .3625$. 

In, Figure (\ref{fig:figP2}), the results of autosynchronization after 2000 seconds of simulation are shown where reconstructed parameters are compared with their true counterparts. Both plots demonstrate how effectively the  parameters have been reconstructed. Similar results were found by testing parameters that vary spatially according to Eq \eqref{eq:Linear}. 


\section{Syncronization by Sampling Only One Species}
To this point, an important criticism of our work is that we need to sample both species to drive the response model and parameters. As mentioned above, our interest in autosynchronization for parameter estimation stems from work with ocean models for phytoplankton-zooplankton ecology. In fact, hyperspectral satellite imagery provides phytoplankton density inferences but provides no data for zooplankton. Certainly, parameter estimation using the response model above will fail. Even given correct model parameters, it is impossible to forecast the model since zooplankton initial conditions are not supplied. Our problem of interest requires that we somehow estimate zooplankton initial conditions based on phytoplankton observations. 

\begin{figure}[h]
\begin{center}
\subfigure{\includegraphics[width=3.2in]{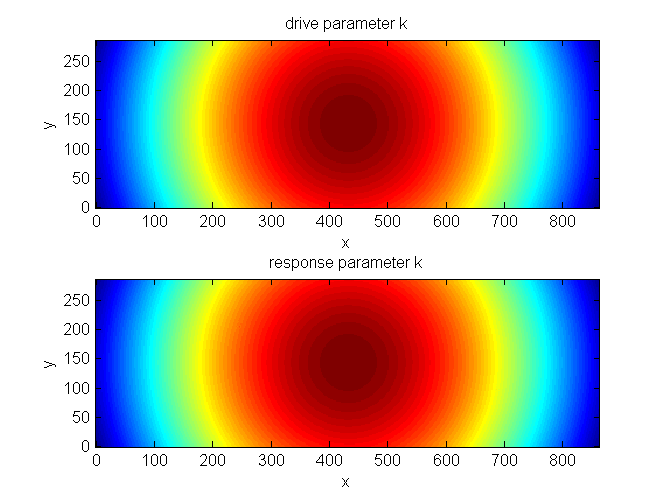}}
\subfigure{\includegraphics[width=3.2in]{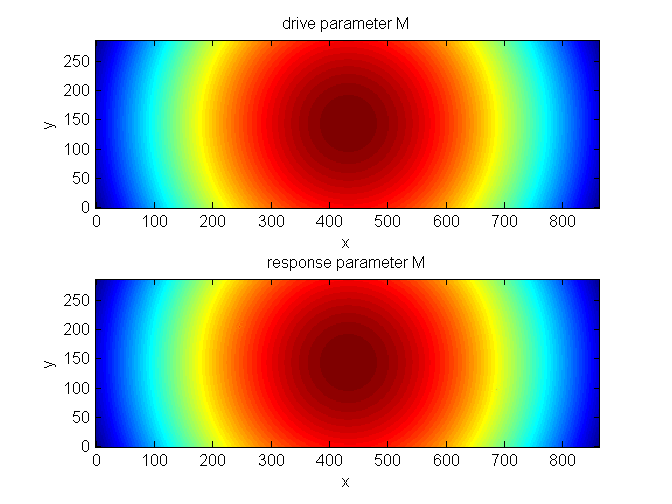}}
\caption{Autosynchronization of spatially dependent parameters at $t=4434$. Pictured as drive (top) vs response (bottom) with $k_2(x,y)$ vs $\hat{k}(x,y,4434)$ on the left and $m_2(x,y)$ vs $\hat{m}(x,y,4434)$ on the right.}
\label{figParkmPhytoOnly}
\end{center}
\end{figure}

We find that, by a modification of Eq \eqref{eq:Fishr}, it is possible to drive zooplankton density to its true state by sampling phytoplankton alone. This is a first demonstration of the possibility of simulating this system with only partial knowledge. As an added bonus we observe autosynchronization. Thus this technique gives us a tool to estimate parameters and to initialize a model for short term forecasts. The response model that drives these results is

\begin{eqnarray}\label{eq:FishPhyto}
\frac{\partial \hat{P}}{\partial t} &=& \triangle \hat{P} + \hat{P}(1-\hat{P}) - \frac{\hat{P} \hat{Z}}{\hat{P} + h} + \kappa(P-\hat{P}),\\
\nonumber \frac{\partial\hat{Z}}{\partial t} &=& \triangle \hat{Z} + \hat{k}\frac{\hat{P} \hat{Z}}{P+h} - \hat{m}\hat{Z},
\end{eqnarray}
where the absence of hats denotes where observation data is coupled directly into the PDE. We use a combination of diffusive coupling and complete replacement coupling in the response PDE to observe autosynchronization. Note  that zooplankton density is no longer observed in Eq \eqref{eq:FishPhyto}. The  parameter update equations are 
\begin{eqnarray}\label{eq:ParUpdatePhytoOnly}
\frac{\partial \hat{k}}{\partial t} &=& s_1(P - \hat{P}) \\
\nonumber \frac{\partial\hat{m}}{\partial t} &=& s_2(P - \hat{P})\hat{P},
\end{eqnarray} 
with $s_1 = 0.2$, $s_2 = 0.6$, and $\kappa = 0.6$. Here, as above, the parameter equations are evolved simultaneously with the \eqref{eq:FishPhyto} using a forward Euler discretization. Figure (\ref{figParkmPhytoOnly}) shows results obtained when autosynchronizing with Eq \eqref{eq:FishPhyto} at $t = 4434$, a substantially longer time epoch. Results obtained for the Gaussian form, Eq \eqref{eq:Gauss}, for parameters show that both $\hat{k}(x,y,t)$ and $\hat{m}(x,y,t)$  converge, as before, to their true values $k_2(x,y)$ and $m_2(x,y)$. Here, all initial conditions are set to $\hat{P}(x,y,0) = \hat{Z}(x,y,0) = \hat{k}(x,y,0) = \hat{m}(x,y,0) = 1$. 

\begin{figure}[h!]
\begin{center}
\subfigure{\includegraphics[width=3.2in]{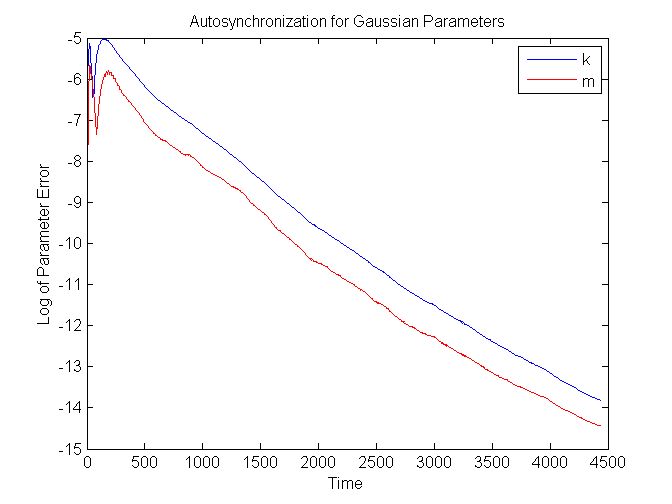}}
\subfigure{\includegraphics[width=3.2in]{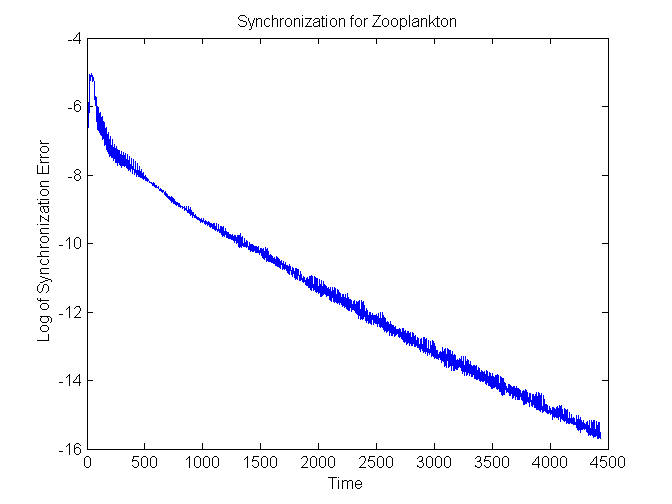}}
\caption{Left: Globally averaged synchronization error between drive and response parameters on a log scale. Right: Globally averaged synchronization error between drive and response PDEs (Zooplankton) on a log scale. Both results are obtained using Eq \eqref{eq:FishPhyto} with drive model parameters $k_2(x,y)$ and $m_2(x,y)$.}
\label{figErrorsPhytoOnly}
\end{center}
\end{figure}

In Figure (\ref{figErrorsPhytoOnly}), globally averaged errors are shown to diminish over time as the synchronization scheme evolves. Both parameters converge to within about $1.0 \times 10^{-6}$ after 4434 seconds. Importantly, we note zooplankton convergence to nearly $1.0 \times 10^{-7}$ of ground truth. Therefore, we need not sample zooplankton to observe autosynchronization and we find the true zooplankton density profile such that model simulations may be initialized.

\section{Conclusion}
In this paper, we have shown that it is possible to derive an autosynchronization scheme for a system of PDEs. We emphasize here the improvements we have made upon past synchronization methods in that we use autosynchronization as a means of parameter estimation of parameters that exist in a function space. We assume that we know the model form of the true observed system, but have no prior knowledge of the parameters. By sampling at every time step, we observed identical synchronization between the response and drive systems as described in \cite{BLP11}. As a first attempt, we have given a model form for adaptive parameters in the response system such that we observe identical synchronization between response model parameters and true parameters, or autosynchronization. Our techniques were implemented on a benchmark model and results converge to ground truth. Thus, autosynchronization is observed for PDEs with spatially homogeneous parameters.

Next, we considered the same system of PDEs wherein the parameters were spatially dependent. We provided a scheme with which we observe autosychronization of spatially dependent parameters. We tested our results against several different functional forms for parameters and found the method to be robust.  

We markedly improved upon these results once more with an autosynchronization scheme that requires sampling of only one species (phytoplankton). We noted that in order to evolve a system of PDEs for forecasting, we need initial conditions for both species; this is a serious problem when dealing with remote sensing data with which we can only observe one of the species. This concern was addressed by providing a response system that autosynchronizes parameters and synchronizes zooplankton using only phytoplankton data. These methods are therefore plausible for use in remote sensing problems. 

As discussed above, synchronization schemes can be proven to work for a given range of coupling parameters using, for example, Lyapunov functions. It remains to be shown why this scheme works on this system, and to perhaps derive autosynchronization model forms for a wider class of reaction-diffusion PDEs. 

A drawback of this technique with application to hyperspectral satellite data is that data may be noisy; this is where filtering techniques have a built-in advantage. Data may also be occluded because of cloud cover. Since there is no hope to synchronize PDEs without data, we require techniques to fill in that which is missing, for example, inpainting. 
 
Another problem with applying these techniques to satellite imagery is temporal data resolution. There may be several days between images and autosynchronization requires ample data observations. The need for frequent observables is perhaps the main drawback to this method. However, autosynchronization may be advantageous for parameter estimation or model building in situations where spatiotemporal data are abundant and especially where parameters are expected to vary spatially.

\section{Acknowledgements}
This work was supported by the Office of Naval Research under grant \#N00014-09-1-0647. The authors would also like to the thank the anonymous referee for helpful comments and suggestions on the manuscript.

\bibliographystyle{spmpsci}
\bibliography{AutoSync}
  
\end{document}